\documentclass[a4paper,12pt,english]{article}

\usepackage{amsfonts,bm,amssymb,euscript,array,babel,cite}

\usepackage{amsmath,amsthm}
\usepackage{pictexwd,dcpic}
\usepackage[dvips]{graphicx}
\usepackage{slashed}

\makeatletter

\newcommand{\f}[2]{\frac{#1}{#2}}

\newcommand{\dd}{\mathrm{d}}
\newcommand{\e}{\mathrm{e}}
\newcommand{\w}{\wedge}
\newcommand{\bbm}{\left(\begin{matrix}}
\newcommand{\ebm}{\end{matrix}\right)}
\newcommand{\beq}{\begin{eqnarray}}
\newcommand{\eeq}{\end{eqnarray}}
\newcommand{\tr}{\text{tr}}
\makeatother

\newcommand{\tdual}[1]{\overset{#1}{\longmapsto}}
\newcommand{\sdual}{\overset{S}{\longmapsto}}

\newcommand{\sfrac}[2]{{\textstyle\frac{#1}{#2}}}

\newcommand{\be}{\begin{equation}}
\newcommand{\ee}{\end{equation}}

\newcommand{\beqa}{\begin{eqnarray}}
\newcommand{\eeqa}{\end{eqnarray}} 
\def\nn{\nonumber} \def \bea{\begin{eqnarray}} \def\eea{\end{eqnarray}}

\newcommand{\barr}{\begin{array}}
\newcommand{\earr}{\end{array}}
\numberwithin{equation}{section}

 \def\g{\gamma} 
 \def\d{\delta} 
\def\f{\phi}    
\def\l{\lambda} \def\L{\Lambda}  \def\m{\mu}
\def\n{\nu}    
\def\s{\sigma}  \def\t{\tau}

\def\Z{{\mathbb Z}} \def\one{\mbox{1 \kern-.59em {\rm l}}}

\def\bit{\begin{itemize}} \def\eit{\end{itemize}}

\def\({\left(} \def\){\right)}

 \allowdisplaybreaks[3]

\begin{document}

\makeatother


\parindent=0cm

\renewcommand{\title}[1]{\vspace{10mm}\noindent{\Large{\bf

#1}}\vspace{8mm}} \newcommand{\authors}[1]{\noindent{\large

#1}\vspace{5mm}} \newcommand{\address}[1]{{\itshape #1\vspace{2mm}}}


\begin{titlepage}
\begin{flushright}
\small ITP-UH-17/13\\
\normalsize
\end{flushright}
\vspace{1cm}

\begin{center}
\title{ {\Large Effective actions of non-geometric fivebranes}}

\vskip 3mm

\authors{\normalsize Athanasios {Chatzistavrakidis${}^{1}$}, Fridrik Freyr Gautason${}^{1,2}$,\\ George 
{Moutsopoulos${}^{3}$}}
and Marco Zagermann${}^{1,2,3}$

\vskip 3mm

\address{  {${}^1$}Institut f\"ur Theoretische Physik,

\&

{${}^2$}Center for Quantum Engineering and Spacetime Research

\&

{${}^3$}Riemann Center for Geometry and Physics, \\ Leibniz Universit\"at Hannover, Appelstra{\ss}e 2, 
30167 Hannover, Germany}

\bigskip 
\small E-mails: thanasis@itp.uni-hannover.de, fridrik.gautason@itp.uni-hannover.de, 
Marco.Zagermann@itp.uni-hannover.de, gmoutso@gmail.com
%
%

\vskip 1.4cm

\textbf{Abstract}

\vskip 3mm

\begin{minipage}{14cm}%

An interesting consequence of string dualities is that they reveal situations where the geometry of a string background appears to be globally ill-defined, 
a phenomenon usually referred to as non-geometry. 
On the other hand, string theory contains extended objects with non-trivial monodromy around them, often dubbed 
defect or exotic branes in co-dimension two.
We determine and examine the worldvolume actions and the 
couplings of certain such branes. In particular, based on specific chains of 
T and S-dualities, we derive the DBI and WZ actions, which describe
 the dynamics of type IIB  fivebranes as well as their couplings 
 to the appropriate gauge potentials associated to mixed symmetry tensors. Based on these actions we discuss how these 
branes act as sources of non-geometric fluxes. In one case this flux is what is usually termed Q flux, associated to 
a T-fold compactification, while in the S-dual case 
 a type of non-geometry related to the Ramond-Ramond sector is encountered.

\end{minipage}

\end{center}

\end{titlepage}

\tableofcontents

\section{Introduction}\label{intro}


String theory contains a plethora of extended objects apart from 
fundamental strings. The most important of them are 
D$p$-branes, which are (1+p)-dimensional hypersurfaces charged under 
the Ramond-Ramond (RR) gauge potentials of type II superstring theories\cite{Polchinski:1995mt}. Moreover, they are dynamical 
objects with tension inversely proportional to the string 
coupling and hence provide an excellent window 
to non-perturbative aspects of string theory. Apart from 
D$p$-branes, the type II superstring theories (as well as the 
heterotic strings) contain Neveu-Schwarz 5-branes (NS5).
 These extended objects 
couple magnetically to the NSNS gauge potential and as such they 
are the magnetic cousins of Fundamental  strings (F1). 
At weak string coupling they are heavier than D$p$-branes, in 
the sense that their tension is inversely proportional 
to the second power of the string coupling. Other standard extended objects
are the Kaluza-Klein Monopoles (KKM), which couple to a 
Kaluza-Klein gauge field. All these objects were studied 
extensively over the past years.

The second wide window to non-perturbative string physics was 
opened by the discovery of string dualities \cite{hulltownsend}. 
Moreover, it was soon realized that the interplay between dualities 
and branes leads to a whole new family of extended objects, 
with D$p$, NS5 and KKM being just a fraction of it \cite{exoticelitzur,exoticblau,exotichull,meessenortin}.
These new BPS-objects have codimension 2 and are obtained as orbits
 of the U-duality group. 
Some of them are even heavier than the NS5-brane, 
exhibiting a tension inversely proportional to the third or even 
fourth power of the string coupling. For a long time these branes were not much explored and remained in the shadow of more standard branes. 
Recent studies, however, suggest that such exotic branes 
have interesting properties and deserve more attention \cite{exotic,defect,axel,Kikuchi:2012za,Hassler,Kimura}. The property of non-standard branes 
that we will be mainly concerned with in this paper is that 
they induce non-trivial monodromies around them that
generate what is often termed non-geometry\footnote{See also \cite{McOrist:2010jw} for a discussion of 5-brane sources with nonstandard monodromies in the heterotic string.} \cite{exotic}. 

Non-geometry is another property of string theory whose origin 
is based on dualities. It was realized many years ago that 
performing T-dualities on known string backgrounds often 
leads to backgrounds which are not globally well-defined in terms of conventional geometric quantities\cite{kstt}. 
A description of the full current status on the topic is beyond 
the scope of this paper. Here we would like to mention that 
one of the popular approaches to a better understanding of
such setups is based on an attempt to 
geometrize the apparent non-geometric background in terms 
of a higher-dimensional geometry \cite{hullng}. This method relies 
on the so-called doubled formalism, where an additional set 
of coordinates dual to the standard geometric ones is introduced. 
Such dual coordinates may be thought of as auxilliary (as in the 
case of twisted doubled tori \cite{hulltdt,prezastdt}) or as fundamental, dynamical 
ones (as in double field theory \cite{dft1,dft2}
{\footnote{For some recent reviews and a more complete 
list of references, we refer to Refs. \cite{
dftrev1,dftrev2}.}). 
Another approach to the issue of non-geometry 
uses modern techniques inspired by generalized complex geometry \cite{Gualtieri}, where the structure group of the combined tangent and cotangent bundle coincides with the T-duality
group. 
The set of geometric operations on the NSNS field content of the background, i.e. 
diffeomorphisms and gauge $B$-transformations, is extended to include $\beta$-transformations, 
associated to an antisymmetric 2-vector field which naturally appears in this 
extended formulation \cite{GMPW}. Simply stated, this essentially results in the implementation of T-dualities as 
geometric operations on the fields when they are patched after traversing a non-contractible loop in target space. 

In the present paper we study some aspects of exotic fivebranes, 
in particular their dynamics, their couplings and their 
relations to non-geometry by deriving their effective worldvolume actions. The effective actions of D$p$-branes are the well-known Dirac-Born-Infeld (DBI) action\cite{Callan:1986bc,dbi} and the Wess-Zumino (WZ) action, which describe, respectively, the response of the brane to the NSNS- and the RR-sector of a supergravity background. Moreover, it implements 
T-duality in the sense that the T-dual action of a D$p$-brane
 indeed gives the same action for a D($p\pm$1) brane, 
depending on the direction that T-duality is performed. 
Similar worldvolume actions exist for NS5-branes and KKMs 
\cite{massivebranes,lozano1}. Moreover, the authors of Ref. \cite{lozano2} determined 
the worldvolume actions for some exotic 6- and 7-branes. 
In the present paper, following an approach similar to Ref. \cite{lozano1}, we 
determine the actions for the two exotic fivebranes of the type IIB 
superstring. The first 
is related via two T-dualities to the type IIB NS5-brane\footnote{Note that the NS5-brane breaks the circle isometry relevant for T-duality unless it is smeared \cite{Tong:2002rq}. Nevertheless T-duality to the KKM remains valid also for the localized NS5-brane due to world sheet instanton corrections that lead to a localization of the KKM in winding space \cite{Harvey:2005ab}. Further understanding emerges from the treatment in double field theory \cite{Jensen:2011jna}, also for analogous issue regarding the $5^2_2$-brane \cite{Kimura}.}
 (or, equivalently, via one T-duality to the type IIA 
KKM) and denoted{\footnote{The notation is explained in section \ref{branesreview}.}} as $5^2_2$,
while the second is obtained from the first upon S-duality and denoted as $5_3^2$. 

Although the worldvolume DBI actions may be determined in a rather 
straightforward way, the issue of  determining the WZ
couplings in the case of exotic branes is less trivial. 
These couplings are well-known for D$p$-branes and were derived in \cite{lozano1} and \cite{exotic} for NS5-branes.  
We will see that it is worth revisiting the NS5 couplings, and we 
will express them in a very compact  
form. More importantly, using the appropriate duality rules, we will 
determine the gauge potentials which couple to the exotic fivebranes. 
It turns out that these potentials are magnetic duals of the Kalb-Ramond field (for the $5^2_2$-brane) and the 
RR 2-form (for the $5^2_3$-brane), albeit not the standard magnetic duals to which the D5- and NS5-branes couple. Instead they can naturally be interpreted as higher rank 
forms that also carry vector indices. The role of exotic duals to standard gauge potentials 
in the study of non-standard branes
was pointed out already in Ref. \cite{defect} from an alternative point of view. 
Our results support and clarify this role.

Having written down the worldvolume  actions for the 
two exotic fivebranes, mutually related by S-duality, we 
discuss and clarify their relation to non-geometry\cite{Hassler}. In particular we 
argue that the first brane acts as a source for non-geometric 
Q flux, as expected from its interpretation as a T-fold in the supergravity picture. 
This result is based on a rewriting of the type IIB action 
in terms of suitable variables, similar to the one which was performed in Ref. \cite{andriot}. 
This rewriting allows one to consider the bulk and worldvolume actions on equal 
footing and write down the corresponding modified Bianchi identity\cite{Hassler}.
Another very interesting result is obtained for the second 
fivebrane, which is related to the previous one by S-duality. 
This brane acts as a source of what appears as some sort of non-geometric 
RR flux, a situation which has not been widely discussed in the 
literature (see however Refs. \cite{aldazabalRR1,rrng}). We discuss this RR 
non-geometry and argue that it is a very reasonable outcome, 
which has been slightly overlooked due to focus on T-duality 
and not S-duality which exchanges NSNS with RR potentials. 

The paper is organised as follows. In section \ref{branesreview} we review the fivebranes of type IIB string theory and how they are related by string dualities. In section \ref{NS5brane} we revisit the worldvolume actions for the NS5-brane. In section \ref{exoticDBI} we derive the DBI action for two exotic fivebranes, and in section \ref{exoticWZ} we derive their WZ action and discuss them as sources of non-geometric flux. In section \ref{summary} we summarize the results of this paper. 
 
\section{Preliminaries on dualities and branes}\label{branesreview}

As mentioned in the introduction, string theory contains a plethora of extended objects apart 
from fundamental strings. Let us focus on the type II superstring theories. These contain 
a variety of D$p$-branes, with $p=0,2,4,6,8$ for the type IIA and $p=-1,1,3,5,7$ for the type IIB {\footnote{
Let us recall that the D8-brane is special and related to a massive deformation of type IIA supergravity. 
Moreover, the D7-brane is also special since it is not asymptotically flat\cite{Bergshoeff:1996ui}.
}}. 
D$p$-branes couple to the corresponding RR gauge
 potentials of these theories. In addition, both theories 
contain NS5-branes, which couple magnetically to the Kalb-Ramond 
field of the common sector of the two theories,
as well as KK monopoles, which couple to abelian KK
gauge fields arising from dimensional reduction.

The type II theories are exchanged under the action of T-duality, whereas the type IIB theory also features a strong-weak 
self-duality (S-duality). These stringy symmetries 
lead to new extended states, which were discovered and classified 
in Refs. \cite{exoticblau,exoticelitzur,exotichull}. 
These were revisited recently in Ref. \cite{exotic,Hassler}, where 
it was shown that they are related to  non-geometric 
backgrounds. In this paper, we study the dynamics and supergravity couplings
 of such branes by constructing their effective action and further clarify 
their connections to non-geometry. 

While a generic treatment of all these states would certainly be very 
interesting, it might also obscure some of the details and the
differences between the different types of branes. We will therefore 
make two restrictions in this paper. 
We focus (a) on the type IIB superstring theory, mainly 
because it also involves S-duality which will prove to be very 
interesting and leads to cases not much studied in the literature,
 and (b) on fivebranes 
of this theory. It will turn out that these branes are 
sufficient to study the properties that we are interested in 
and to reach a self-contained set of results.

The fivebrane states of the type IIB superstring are the 
D5-brane ($5_1$), the NS5-brane ($5_2$), the KKM ($5_2^1$)
and two exotic states, the 
$5_2^2$-brane and the $5^2_3$-brane. We use the notation of Ref. 
\cite{exotic}, where the main number denotes the amount 
of worldvolume directions (which is always 5 in this paper, 
unless otherwise stated), the lower index denotes the 
power of the inverse string coupling in the tension of the 
object and the upper index denotes the number of ``special 
(NUT) transverse directions", on which the mass of the 
brane depends quadratically. Therefore we observe that this 
set of objects contains a lot of diversity both in 
their ``non-perturbativity" (the power of inverse string 
coupling is 1,2 or 3){\footnote{It should be mentioned 
that there are also branes with power 4.}} and in the 
amount of special transverse directions (0, 1 or 2){\footnote{More special directions, up to seven, appear 
as one considers lower dimensions.}}.

Let us now discuss how these branes are linked to each 
other by 
dualities. We begin with the D5- and the NS5-brane, which transform into each other under S-duality, as they form magnetic sources for the $SL(2;\Z)$ doublet of the RR-potential $C_2$ and the NSNS potential $B$.
 Next, the NS5-brane is linked to the KKM 
by two T-dualities, one along a transverse and one along 
 a worldvolume direction (the former leads to the type 
IIA KKM, which T-dualizes to the type IIB KKM under the latter).
Moreover, under two transverse T-dualities the NS5 is linked to 
the $5_2^2$-brane. The latter can also be obtained from the KKM 
under two T-dualities, one along a worldvolume and one along 
a transverse direction (the intermediate type IIA brane 
in this case is also an exotic $5_2^2$ state). Finally, 
the $5_3^2$-brane is obtained from the type IIB $5_2^2$ upon S-duality. 
For completeness, let us mention that the IIB KKM is self-dual 
under S-duality. What we explained with words is 
depicted in the accompanying Figure 1.

\begin{figure}[h]
\begin{center}
\includegraphics[scale=0.75]{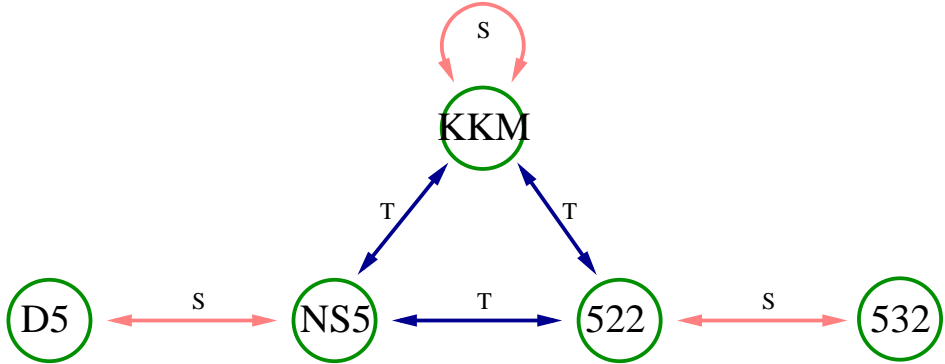}
\end{center}
\caption{\label{fivebranes} The duality chain of fivebranes in type IIB string theory. In the figure S denotes S-duality and T$^2$ denotes two T-dualities along the two directions of a torus.}
\end{figure}

Since we plan to discuss connections of the exotic branes to 
non-geometry, it is suggestive to make contact with the 
corresponding flux chain. The standard T-duality chain of 
fluxes can be written as
\be \label{standardchain}
H_{abc} \overset{T_a}\longleftrightarrow f^a_{\ bc} \overset{T_b}\longleftrightarrow Q^{ab}_{\ \ c} 
\overset{T_c}\longleftrightarrow R^{abc},
\ee
where the leftmost entry refers to a NSNS 3-form flux, which T-dualizes 
to a geometric flux (nilmanifold), while the other two entries, 
are commonly called non-geometric fluxes. In relation 
to the branes that we discuss here, the corresponding chain 
of fluxes reads as follows, where some of the T-dualities are also along directions without flux,
\[\begindc{\commdiag}[200]
\obj(1,0)[f3]{$F_{abc}$}
\obj(4,0)[hh]{$H_{abc}$}
\obj(6,3)[ff]{$f^a_{\ bc}$}
\obj(8,0)[qq]{$Q^{ab}_{\ \ c}$}
\obj(11,0)[pp]{$P^{ab}_{\ \ c}$}
\mor{f3}{hh}{S}
\mor{hh}{ff}{T$^2$}
\mor{hh}{qq}{T$^2$}
\mor{ff}{qq}{T$^2$}
\mor{qq}{pp}{S}
\enddc\]

Some remarks are in order here. First, the chain we 
discuss involves also S-dualities and therefore RR fluxes 
appear as well, in particular $F_3$. This is not common 
in most discussions of non-geometric fluxes but it is important here 
because the rightmost entry of the chain can be viewed as
 a non-geometric RR flux, which we denoted by $P$ \cite{aldazabalRR1,rrng}. 
We will be more specific on this point later on. Furthermore, 
it should be noted that the chain related to the IIB fivebranes 
does not contain an entry associated to the so-called $R$ 
flux. We will have nothing more to say on this point, apart from 
some comments in the final discussion.  

It is also instructive to remember the supergravity 
solutions describing the above standard and non-standard branes.
This will assist our 
discussion in section \ref{exoticWZ}. The first three, D5, NS5 
and KKM, may be found in standard textbooks (see e.g. 
Ref. \cite{blt}, chapter 18).
The solutions for $5_2^2$ and $5_3^2$ were written down in Ref. \cite{gkkm} 
as generalized KKMs.
 These data appear in the
 following table{\footnote{The last two rows of this table should be considered with some caution. 
 They are local expressions which do not make sense globally in standard supergravity formulation. 
 More details on this will be given in section \ref{exoticWZ}.}}.

\begin{center}
 \begin{tabular}{l*{4}{c}r}
 IIB Brane & Metric $\dd s^2$ (string frame)     & Dilaton $e^{2\phi}$ & Flux
 \\
\hline \\
D5 ($5_1$)&  
 $H^{-\sfrac{1}{2}}\dd x^2_{\parallel}+H^{\sfrac 12}\dd x_{\perp}^2$ 
& $H^{-1}$    & 
 $F_3$ 
\\ \\
NS5 ($5_2$)&    $\dd x_{\parallel}^2+H\dd x_{\perp}^2$     &  
$H$ & $H_3$
   \\ \\
KKM ($5_2^1$) &   $\dd x_{\parallel}^2+H\dd x_{\perp}^2+H^{-1}\dd x^2_{\odot}$  
& $1$ &  
$f_2^1$
\\ \\
$5^2_2$&  $\dd x_{\parallel}^2+H\dd x_{\perp}^2+HK^{-1}\dd x_{\odot}^2$  & 
$HK^{-1}$ &  $Q_1^2$ 
\\ \\
$5^2_3$& $(HK^{-1})^{-\sfrac{1}{2}}\dd x^2_{\parallel}+(HK^{-1})^{\sfrac 12}(\dd x_{\perp}^2+\dd x_{\odot}^2)$
& $(HK^{-1})^{-1}$ 
& $P_1^2$ 
\\ \\
\end{tabular} \\ 
\end{center}

In obvious notation, $x_{\parallel}$ denotes the worldvolume directions 
of the brane solution, while $x_{\perp}$ are the ordinary transverse directions. Moreover, 
$x_{\odot}$ denote special transverse directions. 
$H$ is a harmonic function, different for each solution. Its explicit form may be found 
in the textbook \cite{blt}. Furthermore, $K$ appearing in the $5^2_2$ and $5_3^2$ solutions, is 
another function of the form
$$K=H^2+\biggl(\frac{R_{\odot}^2}{2\pi\alpha'}\theta\biggl)^2.$$
Note that these two branes are of co-dimension two\footnote{With co-dimension we mean the number of the ordinary transverse directions $x_\perp$}, with $\theta$ being the polar angle of the transverse directions $x_\perp$. 
Regarding the flux column, the first three rows are obvious. 
The fact that we associated a $Q_1^2$ flux to the fourth row 
was recently discussed to some extent in Refs. \cite{geissbuhler2,Hassler} and will 
be revisited below. The $P_1^2$ in the fifth row is currently 
just a label but it will  be explained 
in section \ref{exoticWZ}.

 The worldvolume actions of all but the exotic fivebranes
 were previously studied in \cite{lozano1}.

\section{The NS5-brane}\label{NS5brane}
In this section we revisit the worldvolume actions for the type IIB NS5-brane which were derived in\cite{lozano1} (see also \cite{Bandos:2000az} for a different approach to the type IIA NS5-brane). For the reader's convenience and to fix our notation (see appendix \ref{notation}), we will rederive these actions.

Let us begin our discussion with the well-known DBI action of the D5-brane
\cite{dbi},
\beq
S_{\text{DBI},\text{D}5} = -T_{\text{D5}} \int_{{\cal M}_{\text{D5}}} \dd^6\sigma\ \e^{-\phi}\sqrt{-\det{(G_{ij}+B_{ij}+\check F_{ij})}},
\eeq
where the indices $i,j$ run along the six
 directions of the  worldvolume ${{\cal M}_{\text{D5}}}$. The physical tension in string frame of the D5-brane 
 is 
$$
\tau_{\text{D5}}=T_{\text{D5}}g_s^{-1},
$$
being inversely proportional to the string coupling 
as for any D$p$-brane.
Here and in the following we set $2\pi \alpha'=1$ since this dimensionful factor (with dimensions of (mass)$^{-2}$) 
can always be reinserted by dimensional analysis. 
Moreover, let us recall that
\[
G_{ij} = \partial_iX^M\partial_jX^N G_{MN}\quad\text{and}\quad
B_{ij} = \partial_iX^M\partial_jX^N B_{MN}
\]
are the pullbacks of the spacetime fields $G$ and $B$ by
 the 10D embedding coordinates $X^M$ of the brane. Finally, 
\be \check F_2=\dd \check {\cal A}_1, \ee 
is the field strength of the worldvolume abelian gauge field living on the brane.

The D5-brane is also charged under the RR fields, in particular, it couples electrically 
to the RR gauge potential $C_6$ (or magnetically to the RR 
gauge potential $C_2$). Its couplings to $C_6$ and lower degree forms 
are determined by the Wess-Zumino action
\be\label{wzd5}
S_{\text{WZ,D5}}=\mu_{\text{D5}}\int_{{\cal M}_{\text{D5}}} C_6-\check{\cal F}_2\wedge C_4+\sfrac 12 \check{\cal F}_2\wedge \check{\cal F}_2\wedge C_2 - \sfrac 16 \check{\cal F}_2
 \wedge \check{\cal F}_2\wedge \check{\cal F}_2C_0=
\mu_{\text{D5}}\int_{{\cal M}_{\text{D5}}} \e^{-\check{\cal F}}{\cal C}|_6,
\ee 
where the bulk fields appearing in brane actions should always be interpreted as pullbacks of the respective 10D fields to the brane worldvolume (here denoted by ${\cal M}_{\text{D5}}$). In the last equality the polyform
$${\cal C}=\sum_{\text{even i}} C_i$$
was used, and the notation $|_6$ indicates that the 6-form part of the expression is to be considered. $\mu_{\text{D5}}$ is the charge of the brane under $C_6$.
This action is obtained as the gauge invariant completion of the coupling to the RR potential $C_6$. In particular, 
$\check{\cal F}_2$ is defined as
\be \check{\cal F}_2=B+\dd \check {\cal A}_1, \ee 
with the gauge transformation rules
\be 
\d B = \dd \L_1,\qquad \d \check {\cal A}_1=-\L_1 \quad\Rightarrow\quad \d\check{\cal F}_2=0.
\ee
Which makes the action gauge invariant under the Kalb-Ramond gauge transformation.

As we already mentioned, the D5- and NS5-branes are S-dual to one another. The relevant S-duality rules can be simply expressed as
\beq\label{Sduality}
\tau \sdual -\frac{1}{\tau},\quad C_2  \sdual
B,\quad B \sdual -C_2\quad\text{and}\quad 
G  \sdual |{\tau}| G,
\eeq
where $\tau = C_0 + i\e^{-\phi}$ is the type IIB axio-dilaton. 

According to the above duality rules, the NS5 DBI action is
\beq
S_{\text{DBI,NS}5} = -T_{\text{NS5}} \int_{{\cal M}_{\text{NS5}}} \dd^6\sigma\e^{-\phi}|{\tau}|\sqrt{-\det\biggl(G_{ij}-
|{\tau}|^{-1}
{{\cal F}}_{ij}\biggl)}
\eeq
with
\be \label{tildef}
{\cal F}_2=C_2+\dd{{\cal A}}_1,
\ee 
where $C_{ij}$ is the pullback of the corresponding
 supergravity potential and ${\cal A}_1$ is the 
worldvolume gauge field, defined such that ${\cal F}_2$ is gauge invariant. Under S-duality $\check{\cal F}_2$  transforms as
 $\check{\cal F}_2 \sdual -{\cal F}_2$. 
 Equivalently, we can write
\beq\label{NS5DBI}
S_{\text{DBI,NS}5} = -T_{\text{NS5}} \int \dd^6\sigma\e^{-2\phi}\sqrt{1+e^{2\phi}C_0^2}\sqrt{-\det\biggl(G_{ij}-
\frac{e^{\phi}}{\sqrt{1+e^{2\phi}C_0^2}} {\cal F}_{ij}\biggl)}.
\eeq
The latter expression makes the non-perturbative nature of the NS5-brane more transparent, as the scaling of the physical tension with the inverse square of the  string coupling becomes manifest,
$$ \t_{\text{NS5}}\propto g_s^{-2}. $$

The WZ couplings of the NS5-brane can be found as before by constructing the fully gauge invariant completion of the coupling 
$$\mu_{\text{NS5}}\int_{{\cal M}_{\text{NS5}}} B_6,$$
where $B_6$ is the magnetic dual of the Kalb-Ramond gauge potential (see eq. \eqref{h7}). In order to carry out this task, we need to know the 
 gauge transformation of $B_6$. This is determined in the 
appendix (eq. (\ref{deltaB6})), and we repeat here the result,
\be \d B_6=\dd\L_5+ \dd\l_3\wedge C_2
+\dd\l_1\wedge B\wedge C_2,\nn \ee
where $\L_5$ is a 5-form gauge parameter, and $\lambda_1,\lambda_3$ are the 1- and 3-form gauge parameters associated with the gauge transformations of $C_2$ and $C_4$, respectively.
This dictates the couplings that have to be added in the WZ action in order to render it gauge invariant. 
 This action may be written in polyform notation as before,
 defining the new polyform
\be \label{nspolyform}
{\cal B} =  \frac{C_0}{|\tau|^2} - B - \left(C_4 - C_2\w B\right)+ \left(B_6 - \frac{1}{2}B\w C_2\w C_2\right).
\ee 
We can now write
\be \label{wzns5b}
S_{\text{WZ,NS5}}=\mu_{\text{NS5}}\int_{{\cal M}_{\text{D5}}} e^{-\cal { F}}{\cal B}|_6.
 \ee 
 This action can be also obtained very easily by directly applying the S-duality 
 rules to the action (\ref{wzd5}) (see appendix A). Indeed, ${\cal B}$ is the (negative) 
 S-dual of the polyform ${\cal C}$. This is the approach followed in Ref. \cite{lozano1}. The approach we employed here 
 was used in Ref. \cite{exotic},
 albeit with different conventions for the gauge potentials (see appendix \ref{gaugetrafo}).

\subsection{Modified Bianchi identities.}
The D5- and NS5-branes act as sources for RR and NSNS 3-form field strengths, respectively. 
Indeed, in their presence the corresponding Bianchi identities are modified, similar to what 
happens in standard electrodynamics in the presence of a magnetic source.
For the NS5-brane the relevant terms are the kinetic term for the Kalb-Ramond potential and 
the corresponding leading WZ coupling, i.e.
\bea 
S_{B_6}=-\int_{10} \sfrac 12 \e^{2\phi} H_7\wedge\star\, H_7+\mu_{NS5}\int_{10} B_6\wedge \d_4.
\eea 
The action is written in terms of the dual field strength $H_7$, and the worldvolume term 
is lifted to ten dimensions by a 4-form $\delta_4$ with support on the worldvolume.
The variation with respect to $B_6$ yields the modified Bianchi identity for the NSNS 3-form,
\be \dd H_3= \star j_6^{NS5}, \ee
where $\star j_6^{NS5} = -\mu_{NS5} \d_4$. This shows that the NS5-brane is a localized source of NSNS flux.

\section{DBI action of exotic fivebranes}\label{exoticDBI}
Our aim here is to determine the analog of the DBI action of the $5_2^2$ brane of the type IIB superstring, which is the double T-dual of the NS5 brane along two of its transverse dimensions, and of the $5_3^2$ brane, which is the S-dual of the $5_2^2$, as explained in section \ref{branesreview}. The WZ couplings of these exotic  branes are discussed in the next section. 

\subsection{T-duality rules}
In the following, we consider a $5^2_2$-brane that is obtained by two T-dualities along two transverse directions of an NS5-brane. The worldvolumes of both branes are taken to be along the directions $034567$, and we perform the two T-dualities along the directions $89$, which form a two-torus. The resulting $5_2^2$-brane then has the transverse directions $1289$ denoted by $r\theta yz$, with $yz$ being the special transverse directions analogous to the one special NUT-like direction of the KKM.

We use the following KK ansatz for the metric\footnote{We generally use hatted symbols for quantities that are invariant under T-duality.},
\[
\dd s^2 = \hat{G}_{\mu\nu} \dd x^\mu \dd x^\nu + G_{mn}(\dd x^m + A^m)(\dd x^n + A^n),
\]
where $A^m = A^m_\mu \dd x^\mu$ are the two KK one forms. The indices  $m,n$ run over the directions $yz$ of the compactified two-torus, 
 $\mu,\nu$ run over the rest, and we use capital indices $M,N$ to denote ten-dimensional directions. Likewise we decompose the NSNS 2-form $B$ as
\[
B = \sfrac 12 \left[\hat B_{\mu\nu} + A_\mu^m B_{m\nu}\right]\dd x^\mu\w \dd x^\nu + B_{m\nu}\dd x^m\w \dd x^\nu\notag + \sfrac12 B_{mn} \dd x^m\w \dd x^n.
\]
In the following, it will often be useful to use the non-coordinate basis $(\dd x^\mu, \eta^m)$ with $\eta^m = \dd x^m + A^m$. In this basis the metric takes the form
\beq\label{twodmetric}
\dd s^2 = \hat{G}_{\mu\nu} \dd x^\mu \dd x^\nu + G_{mn}\eta^m\eta^n,
\eeq
so that the determinant of the metric factorizes conveniently as 
\be 
\sqrt{-\text{det}(G_{MN})}=\sqrt{-\text{det}(\hat G_{\mu\nu})}\sqrt{\text{det}(G_{mn})}.
\ee
To rewrite the Kalb-Ramond field in the basis $(\dd x^\mu, \eta^m)$, it is useful to introduce the 1-form
\beq\label{theta}
\theta_{m} = \theta_{m\mu}\dd x^\mu\quad \text{and}\quad \theta_{m\mu} = B_{m\mu}- B_{mn}A^n_\mu,
\eeq
which is motivated by a simple transformation rule under T-duality (see below).
We can now write $B$ as (cf. \cite{ms}),
\beq
B = \sfrac 12 \hat{B}_{\mu\nu}\dd x^\mu \w \dd x^\nu +  \sfrac 12 B_{mn}\eta^m\w\eta^n  + 
 (\eta^m -  \sfrac 12 A^m)\w\theta_m\label{Bfield}.
\eeq
The T-duality rules for the NSNS background fields 
under the combined T-dualities along the directions $y$ and $z$ 
are
\beq
\e^\phi &\tdual{yz}& \frac{\e^\f}{\sqrt{\det(G_{mn}+B_{mn})}},\label{BRdilaton}\\
G_{mn} &\tdual{yz}& \tilde{G}^{mn}=\frac{\det({G_{kl}})}{\det({G_{kl} + B_{kl}})} G^{mn},\label{BRGinternal}\\
A^m_\mu &\tdual{yz}& \theta_{m\mu},\label{BRA}\\
\hat{G}_{\m\n} &\tdual{yz}& \hat{G}_{\m\n},\\
B_{mn}&\tdual{yz}& \tilde{B}^{mn}=\frac{\det(B_{kl})}{\det({G_{kl}+B_{kl}})} (B^{-1})^{mn},\label{BRBinternal}\\
\theta_{m\mu} &\tdual{yz}& A^m_\mu,\label{BRtheta}\\
\hat{B}_{\m\n} &\tdual{yz}& \hat{B}_{\m\n}.\label{BRBhat}
\eeq
This directly implies that the combination
\beq
\sqrt{\det(G_{mn})}\ \e^{-2\phi}
 = \e^{-2\hat\phi}
\eeq
is T-duality invariant.

The T-duality rules for the RR potentials have to be determined as well. To this end, it turns out that 
it is also useful to work  in the $(\dd x^\mu, \eta^m)$ basis. Specifically, we define the following components,
\begin{eqnarray*}
C_{0} &=& \zeta_{0},\\
C_{2} &=& \sfrac12\zeta_{\mu\nu} \dd x^{\mu\nu} + \zeta_{\mu m} \dd x^\mu \w\eta^m + \sfrac12\zeta_{mn}\eta^m\w\eta^n,\\
C_{4} &=& \sfrac{1}{24}\zeta_{\mu\nu\rho\sigma} \dd x^{\mu\nu\rho\sigma} + \sfrac16\zeta_{\mu\nu\rho s} \dd x^{\mu\nu\rho} \w\eta^s 
+ \sfrac14\zeta_{\mu\nu r s} \dd x^{\mu\nu}\w\eta^r\w\eta^s,
\end{eqnarray*}
which are  the components of the RR potentials in this basis. The usual shorthand notation $dx^{i_1i_2...i_p}=
dx^{i_1}\w dx^{i_2}\w ...\w dx^{i_p}$ is hereby employed.
Then the following duality rules for these forms are obtained as,
\begin{eqnarray*}
\zeta_{0} &\tdual{yz}&  \zeta_{yz} - B_{yz}\zeta_{0},\\
\zeta_{mn} &\tdual{yz}& -\epsilon^{mn}\zeta_{0} + \tilde B^{mn}(\zeta_{yz} - B_{yz}\zeta_{0}),\\
\zeta_{\mu n} &\tdual{yz}& \epsilon^{nm}\zeta_{\mu m},\\
\zeta_{\mu\nu} &\tdual{yz}& \zeta_{\mu\nu yz} - \zeta_{\mu\nu}B_{yz}.
\end{eqnarray*}
The completely antisymetric tensor $\epsilon^{mn}$ appearing in these expressions is defined to be such that $\epsilon^{yz} = 1$. 

Finally we give the duality rules of the worldvolume scalars and gauge potentials relevant for the DBI actions. The degrees of freedom of each brane are given in table \ref{fieldcontent}. The T-duality rule is simply
\beq
X^\nu &\tdual{yz}& X^\nu,\\
X^m &\tdual{yz}& \tilde X_m,\\
{\cal A}_1 &\tdual{yz}& \tilde{\cal A}_1,
\eeq
where $\tilde X_m$ are worldvolume scalars along the transverse direction $yz$ and $\tilde{\cal A}_1$ the gauge field of the $5_2^2$-brane.
\begin{table}[h!]
\begin{center}
\begin{tabular}{lcccc}
Brane 	& Worldvolume		 	& Worldvolume 			& Degrees of \\
	& scalars			& gauge fields			& freedom\\
\hline
\rule{0pt}{15pt}D5		& $X^N=(X^\nu,X^n)$		& $\check{\cal A}_1$	& $4+4$\\
NS5		& $X^N=(X^\nu,X^n)$		& ${\cal A}_1$			& $4+4$\\
$5^2_2$	& $X^\nu, \tilde X_m$	& $\tilde{\cal A}_1$	& $2+2+4$\\
$5^2_3$	& $X^\nu, \tilde X_m$	& ${\cal A}'_1$			& $2+2+4$\\
\end{tabular}
\caption{\label{fieldcontent}The degrees of freedom of the branes discussed in this paper. In all cases the worldvolume gauge potential carries 4 physical degrees of freedom and the worldvolume scalars carry 4 physical degrees of freedom representing the four transverse directions of the branes. Note that S-duality does not affect the worldvolume scalars, but does transform the worldvolume gauge fields as required for gauge invariance.}
\end{center}
\end{table}

\subsection{The $5_2^2$ DBI action.}

Using the above T-duality rules, we can calculate the DBI action of the $5^2_2$-brane. The key benefit of having written 
the metric and the forms in the $(\dd x^\mu,\eta^m)$ basis is that the components of the fields transform nicely under T-duality. Furthermore, the pullback of $\eta^m$ transforms directly to a simple set of 1-forms $\tilde\eta_m$, under T-duality,
\[
\eta_M^m\partial_i X^M= \eta_i^m \tdual{yz} \tilde\eta_{im} = \partial_i \tilde X_m + \theta_{m\mu}\partial_i X^\mu,
\]
where $\tilde X_m$ denotes the T-dual of the corresponding worldvolume field.

At this stage the pullback of each field appearing in the NS5 worldvolume action can be easily  transformed. 
The pullback of the metric takes the form
\beq
G_{ij} &=& \hat{G}_{\mu\nu}\partial_i X^\mu\partial_j X^\nu + G_{mn}\eta^m_i\eta^n_j \notag\\
&\tdual{yz}& \hat{G}_{\mu\nu}\partial_i X^\mu\partial_j X^\nu + \tilde G^{mn} \tilde\eta_{im} \tilde\eta_{jn}.\notag
\eeq
Moreover, the pullback of the 2-form is
\beq
C_{ij} &=& \zeta_{\mu\nu}\partial_i X^\mu\partial_j X^\nu + 2\zeta_{\mu n}\partial_{[i} X^\mu \eta^n_{j]}+ 
\zeta_{mn}\eta^m_i\eta^n_j \notag\\
&\tdual{yz}& \left(\zeta_{\mu\nu yz} - \zeta_{\mu\nu}B_{yz}\right)\partial_i X^\mu\partial_j X^\nu +
2\epsilon^{nm}\zeta_{\mu m} \partial_{[i} X^\mu \tilde\eta_{j]n} \notag\\
&&+\biggl( -\epsilon^{mn}\zeta_{0} + \tilde B^{mn}(\zeta_{yz} - B_{yz}\zeta_{0})\biggl)\tilde\eta_{im}\tilde\eta_{jn}.\notag
\eeq
We can now define the gauge invariant 2-form
\beq\label{k2}
\tilde{\cal F}_{ij} &=& 2\partial_{[i} \tilde {\cal A}_{j]} + \left(\zeta_{\mu\nu yz} - \zeta_{\mu\nu}B_{yz}\right)\partial_i X^\mu\partial_j X^\nu +
2\epsilon^{nm}\zeta_{\mu m} \partial_{[i} X^\mu \tilde\eta_{j]n} \notag\\
&&+\biggl( -\epsilon^{mn}\zeta_{0} + \tilde B^{mn}(\zeta_{yz} - B_{yz}\zeta_{0})\biggl)\tilde\eta_{im}\tilde\eta_{jn},\notag
\eeq
which is the T-dual of ${\cal F}_{ij}$, defined in eq. (\ref{tildef}). Finally, the axio-dilaton $\tau$ transforms to the dual modulus 
\[
\tilde\tau = (C_{yz}-B_{yz}C_{0}) + i\sqrt{\det(E_{mn})}\e^{-\phi},
\] 
where $E_{mn}=G_{mn}+B_{mn}$ as usual.
The full DBI action for the $5^2_2$ can now be written down and acquires the form
\beq
S_{\text{DBI},5^2_2} &=& -T_{5^2_2}\int_{{\cal M}_{{5^2_2}}}\dd^6\sigma\ \e^{-\phi}|\tilde\tau|\sqrt{\det(E_{kl})}\notag\\
&& \times\sqrt{-\det(\hat G_{\mu\nu}\partial_iX^\mu \partial_jX^\nu + \tilde G^{mn}\tilde\eta_{im}\tilde\eta_{jn} - |\tilde\tau|^{-1}\tilde{\cal F}_{ij})}.
\eeq
By noticing that $\tilde\tau$ is proportional to $\e^{-\phi}$ we see that the above action has the expected $g_s^{-2}$ dependence.

\subsection{S-duality and the $5_3^2$ DBI action.}

Using the S-duality rules \eqref{Sduality} we can also write down the DBI action for the $5^2_3$-brane. We define 
\bea 
G'^{mn} &=& \frac{\det(|\tau|G_{kl})|\tau|^{-1}G^{mn}}{{\det(|\tau|G_{kl}- C_{kl})}},\nn\\ 
C'^{mn} &=& \frac{\det(C_{kl})}{{\det(|\tau|G_{kl}- C_{kl})}} (C^{-1})^{mn},\label{tildeC}
\eea
which are the symmetric and antisymmetric part, respectively, of the inverse of $|\tau|G_{mn}- C_{mn}$ to which 
$E_{mn}$ transforms.  We then find
\[
\tilde\tau\sdual \tau' = B_{yz}- |\tau|^{-2}C_{yz} C_{0} + i\sqrt{\det(|\tau|G_{mn}-C_{mn})}|\tau|^{-2}\e^{-\phi}.
\]
Furthermore
\beq
\tilde{\cal F}_{ij} &\sdual& {\cal F}'_{ij} = 2\partial_{[i} {\cal A}'_{j]} + \left(\zeta_{\mu\nu yz} -\zeta_{\mu\nu}B_{yz} + 2\epsilon^{mn}\zeta_{\mu m}\theta_{\nu n}\right)\partial_iX^\mu \partial_jX^\nu \notag\\
&&+2\epsilon^{nm}\theta_{\mu m}  \partial_{[i}X^\mu \eta'_{j]n}+
\left[ \epsilon^{mn}|\tau|^{-2} C_{0} + C'^{mn}(B_{yz} - |\tau|^{-2}C_{yz} 
C_{0}\right]\eta'_{im}\eta'_{jn},\nn
\eeq
where 
\[
\eta'_{in} = \partial_i \tilde X_m - \zeta_{m\mu}\partial_i X^\mu.
\]
We can now write down the DBI action for the $5_3^2$
\begin{eqnarray}
S_{\text{DBI},5^2_3} &=& -T_{5_3^2}\int\dd^6\sigma\ \e^{-\phi}|\tau|^{-2}|\tau'|\sqrt{\det(|\tau|G_{kl}-C_{kl})}\notag\\
&& \times\sqrt{-\det(\hat G_{\mu\nu}\partial_iX^\mu \partial_jX^\nu + |\tau|^{-1}\tilde G'^{mn}\eta'_{im}\eta'_{jn} - |\tau'|^{-1}{\cal F}'_{ij})}.
\end{eqnarray}
Here we can again extract the $g_s$ dependence of each factor and see that this action scales as $g_s^{-3}$ as expected.

\section{WZ actions of exotic fivebranes and non-geometry}\label{exoticWZ}

Having discussed the DBI action of the $5_2^2$- and $5_3^2$-branes, we now turn to their WZ actions. The WZ couplings we derive will make the interpretation of these branes as sources for non-geometric fluxes very transparent, complementing the effective supergravity analysis of \cite{exotic,Hassler}. Moreover, they will clarify the meaning of the mixed symmetry forms that occur in group theoretical considerations\cite{defect}.

\subsection{WZ actions of exotic fivebranes}
Writing down the WZ action of the exotic brane involves the T-dualization of $B_6$, the magnetic dual of $B$, that occurs in the WZ action of the NS5-brane. 
Let us recall that $B_6$ is defined by the non-local expression
\beq\label{H7def}
H_7 = \dd B_6 + \cdots = \e^{-2\phi} \star \dd B,
\eeq
where the dots represent the RR terms necessary for the consistent definition of $B_6$. They are given explicitly in the appendix (eq. (\ref{h7})) and lead to the non-trivial Bianchi identity,
\[
\dd H_7 = F_3\w F_5 - F_1\w F_7,
\] 
which means that $H_7$ can be set equal to $\dd B_6$ only under the assumption that the RR fields vanish.
Let us for the moment employ this simplification and neglect the RR terms; i.e. we use $H_7 = \dd B_6$, 
keeping in mind that the expressions below are subject to corrections from the RR sector. 
Starting from the NS5-brane top-form coupling to $B_6$ it is clear that the $5^2_2$-brane couples to a magnetic dual 
of the double T-dual of $B$. We have seen in eq. (\ref{BRBinternal}-\ref{BRBhat}) that the T-dual of $B$ is a 
combination of fields from the NSNS sector, namely the $B$ itself, the KK 1-forms $A^m_\mu$ and the internal metric $G_{mn}$. Since T-duality 
mixes fields from the NSNS sector only, this means that $B_6$ maps under double T-duality to \emph{some}
magnetic dual of the above mentioned fields from the NSNS sector. We will see that this magnetic dual can be naturally described in terms of an 8-form $B_8^2$ with two vector indices (see also \cite{defect}).

\subsection{The KK monopole}
As a warm up, let us begin by considering a single T-duality of $B_6$ which should give the top-form coupling to the type IIA KKM. 
The full WZ action for the KKM was worked out in detail in \cite{massivebranes,lozano1},
but here we are only interested in the top-form coupling\footnote{Higher derivative WZ couplings of D-branes and their T-duals were discussed in \cite{McOrist:2012yc}.}. 
Consider the metric ansatz in eq. \eqref{twodmetric} where the internal indices $m,n,\dots$ only take a single
value, $m=z$, and the $B$ field ansatz (cf. \eqref{Bfield}) is
\[
B = \hat{B}  + \eta^z \w \theta_z - \frac{1}{2}A^z\w \theta_z,
\]
where $\hat{B} = \frac12 \hat{B}_{\mu\nu} \dd x^\mu\w \dd x^\nu$.
Since there is only one internal direction, $z$, the $B_{mn}$ of eq. \eqref{theta} vanishes and $\theta_{z\mu} = B_{z\mu}$.
In order to obtain the magnetic dual $B_6$, we first determine the field strength $H=\dd B$,
\beq\label{dB}
H =  \dd \hat{B} + \frac12 \left(\dd A^z \w \theta_z  + A^z\w \dd \theta_z\right) - \eta^z \w \dd \theta_z =\hat H - \eta^z\w\dd\theta_z,
\eeq
where we have defined the manifestly T-duality invariant 3-form $\hat H$ and used $\dd \eta^z = \dd A^z$. Note that, because of the assumed isometry in the $z$ direction, none of the above differentials on the right hand side of eq. \eqref{dB} have legs in the $\eta^z$ direction. With the current ansatz \eqref{twodmetric} for the metric the Hodge star factorizes and can be written in terms of $\hat{\star}$, the Hodge operator associated with the reduced metric $\hat{G}_{\mu\nu}$ (cf. \eqref{twodmetric}). Then
\beq
\dd B_6 &=& \e^{-2\phi} \star\ \dd B\notag\\
&=& \e^{-2\phi}\sqrt{G_{zz}}(\hat{\star}\ \hat{H})\w \eta^z + \e^{-2\phi}\sqrt{G^{zz}}\hat\star\ \dd\theta_z.\label{dB6}
\eeq
At this stage it is beneficial to define the projection operators
\[
P_z\omega_p = \dd z\w \iota_z\omega_p\quad\text{and}\quad \tilde P_z\omega_p = \eta^z\w\iota_z\omega_p,
\]
where $\omega_p$ is any $p$-form. They are related through $\tilde P_z\omega_p = P_z\omega_p + A^z\w\iota_z\omega_p$.
Here we have defined the contraction with the Killing vector $\partial_z$ by $\iota_z$\footnote{Our conventions are such that if a form $\omega_p$ is decomposed as $\omega_p = \hat\omega_p + \eta^z \w \psi_{p-1}$ then $\iota_z \omega_p = \psi_{p-1}$.}. With the assumed Killing isometry we can make use of the fact that $\dd\iota_z\omega_p + \iota_z \dd\omega_p = {\cal L}_z\omega_p = 0$ to show that $\dd P_z \omega_p = P_z \dd \omega_p$.
Now we can decompose eq. \eqref{dB6}
\beq
(1 - \tilde P_z)\dd B_6 &=& \e^{-2\phi}\sqrt{G^{zz}}\ \hat{\star}\ \dd \theta_z,\label{dB6hat}\\
\dd\iota_z{B}_6 = -\iota_z\dd B_6 &=& -\e^{-2\phi}\sqrt{G_{zz}}\hat{\star}\ \hat{H}.\label{izdB6}
\eeq
We can now easily find the T-dual of $\iota_z\dd B_6$. Using the fact that the combination $\e^{-2\phi}\sqrt{G_{zz}}$ is T-duality invariant, everything on the right hand side in eq. \eqref{izdB6} is T-duality invariant, and we conclude that
\beq
\iota_z B_6 \tdual{z} \iota_z B_6.\label{BRizB6}
\eeq
We are interested in the components of $B_6$ that couple to a NS5-brane transverse to the $z$ coordinate. This is the combination
\[
(1 - P_z)B_6 = (1-\tilde P_z)B_6 + A^z\w \iota_z B_6.
\]
In order to determine the top-form coupling to the KKM we need to T-dualize this, which can be done using \eqref{dB6hat} and \eqref{izdB6}. We calculate
\beq
\dd (1-P_z)B_6 &=& (1-P_z)\dd B_6\nn\\
&=& (1-\tilde P_z)\dd B_6 + A^z \w (\iota_z \dd B_6)\notag\\
&=&\e^{-2\phi}\sqrt{G^{zz}}\ \hat\star\ \dd \theta_z - A^z\w \dd(\iota_z B_6)\notag\\
&\tdual{z}& \e^{-2\phi}\sqrt{G_{zz}^3}\ \hat\star\ \dd A^z - \theta_z\w \dd(\iota_z B_6)\notag\\
&=& \dd( \iota_z A_7^z),\label{beforeT}
\eeq
where we have used the T-duality rules derived before. In the last step we have expressed the result in terms of the magnetic dual of $A^z$. In the nine-dimensional theory this is a 6-form that is guaranteed to be exact by the equations of motion in the reduced theory. This is clear since the reduced theory is $O(1,1)$ invariant and the expression we start with, before performing the T-duality \eqref{beforeT}, is exact. From a ten-dimensional perspective this 6-form can naturally be interpreted as a contraction of a 7-form $A_7^z$, which is the ten-dimensional magnetic dual of the KK 1-form $A^z$. We discuss the reduced theory and magnetic duals in more detail in appendix \ref{appreduced}.
We then get the T-duality rule
\beq
(1-P_z)B_6 \tdual{z}  \iota_z A_7^z,
\eeq
which is valid up to closed forms (and RR-corrections). 
This rule agrees with the results of \cite{lozano1}, although a different but equivalent definition of $A_7^z$ was used.

\subsection{The $5_2^2$-brane}
Let us  now return to the $5^2_2$-brane. 
After having determined the T-duality rules for $B_6$ in the case of a single T-duality, the next step is a straightforward generalization. The $B$ field ansatz becomes \eqref{Bfield}
\[
B = \hat{B}  + \eta^m\w \theta_m - \frac12 A^m\w \theta_m + \frac12 B_{mn}\eta^m\w \eta^n.
\]
The field strength of $B$ is then
\beq
H &=& \hat H - H_m\w \eta^m + \frac12 \dd B_{mn}\w\eta^m\w\eta^n,\label{Hingeneral}
\eeq
where 
\beq
\hat{H} &:=& \dd \hat B + \frac12 \left(\dd A^m\w \theta_m + A^m\w \dd \theta_m\right),\label{hatH}\\
H_m &:=& \dd \theta_m + B_{mn}\dd A^n.\label{Hm}
\eeq
The magnetic dual can be expressed in terms of the fields of eqs. (\ref{hatH},\ref{Hm}) as
\beq
\dd B_6 &=& \e^{-2\phi}\ \star\ \dd B\notag\\
&=&\e^{-2\phi}\sqrt{\det(G_{kl})}\ (\hat{\star}\hat H)\w\eta^y\w\eta^z\notag\\
&& +\e^{-2\phi}\sqrt{\det(G^{kl})}\ \hat{\star} H_m\w \epsilon^{mr}G_{rs}\eta^{s}\notag\\
&& + \e^{-2\phi}\sqrt{\det(G^{kl})}\ \hat\star\ \dd B_{yz}.\notag
\eeq
We immediately see that under double T-duality,
\[
\iota_y\iota_z B_6 \tdual{yz} \iota_y\iota_z B_6.
\] 
Since we start with an NS5-brane that does not wrap the $yz$ directions, the component of $B_6$ that couples to the brane is $(1-P_z)(1-P_y)B_6$. The correponding field strength is
\beq
\dd(1 - P_z)(1 - P_y) B_6 &=& \e^{-2\phi}\sqrt{\det(G_{kl})}\Big[\det(G^{kl})\left(\hat\star \dd B_{yz} + \hat\star H_m\w\epsilon^{mr}G_{rs} A^s\right)\notag\\
&& + \hat\star\hat H \w A^y\w A^z\Big]\notag,
\eeq
which dualizes under double T-duality to
\beq\label{522Tdual}
\e^{-2\phi}\sqrt{\det(G_{kl})}\left[\det(G_{np}+B_{np})\left(\hat\star \dd \tilde B^{yz} - \hat\star \tilde H^m \w \epsilon_{mr}\tilde G^{rs}\theta_s\right) + \hat\star \hat H\w\theta_y\w\theta_z\right],
\eeq
where, as before,
\[
\tilde B^{mn} = \frac{\det(B_{kl})}{\det(G_{kl}+B_{kl})} (B^{-1})^{mn},
\]
and
\[
\tilde H^m = \dd A^m + \tilde B^{mn}\dd \theta_n.
\]
The indices on $\tilde B^{mn}$ are lifted because of its relation to $B^{-1}$. In this sense, $\tilde B$ should not be 
interpreted as a 2-form, but as a 2-vector-valued scalar, with a field strength 
$\dd \tilde B$ which is a 2-vector-valued 1-form. The form that couples to the $5^2_2$-brane in eq. \eqref{522Tdual} is a magnetic dual of this 2-vector-valued scalar which is also 2-vector-valued and has a natural ten-dimensional origin which we call $B_8^2$ for now. The superscript $2$ indicates the number of vector indices. We therefore find the T-duality rule
\beq\label{B6doubleTdual}
(1 - P_z)(1 - P_y) B_6 \tdual{yz} \iota_y\iota_z B_8^{yz},
\eeq
where
\beq
\dd \iota_y\iota_z B_8^{yz} &=& \e^{-2\phi}\sqrt{\det(G_{kl})}\left[\det(G_{np}+B_{np})\left(\hat\star \dd \tilde B^{yz} - \hat\star \tilde H^m \w \epsilon_{mr}\tilde G^{rs}\theta_s\right)\right.\notag\\
&&\left. + \hat\star \hat H\w\theta_y\w\theta_z\right].\label{B82definition}
\eeq
The reason we can be certain that the right hand side of equation \eqref{B82definition} is closed is that it is T-dual to a closed expression. Just as for the KKM, this is a result of the equations of motion in the reduced theory. Since the reduced theory is $O(2,2)$ invariant, the equations of motion must also be satisfied in the dual frame and therefore the derivative of the expression on the right hand side of \eqref{B82definition} must also vanish due to the equations of motion. This may also be verified explicitly using the equations of motion in the reduced theory but the calculation is not very illuminating so we don't perform it here. 
The rule \eqref{B6doubleTdual} determines the top-form coupling to the $5_2^2$-brane.

This result was predicted in \cite{defect} using different methods. Decomposing the adjoint representation of the U-duality
group in terms of representations of the T-duality group the authors of \cite{defect} found more degrees of freedom than 
could be accounted for by just considering the metric, NSNS 2-form, the RR sector and their magnetic duals. 
They concluded that more than one type of magnetic dual exists for almost all fields, and this is precisely what we observe 
here. In \cite{defect} these different types of magnetic duals were explained with the use of {mixed symmetry forms}
\cite{mixed1}.
A mixed symmetry form \cite{mixed1,mixed2,mixed3} (see also Ref. \cite{mixed4} for a related review in a different 
context) is a tensor with two or more sets of antisymmetrized indices. For example a mixed symmetry form
$B_{8,2}$ has at the same time $8$ standard form indices and $2$ \emph{additional} form indices, i.e. the first $8$ indices
of $B_{8,2}$ are completely antisymmetric and the last two are also antisymmetric. According to \cite{defect} the brane 
$5^2_2$ should couple to the form $B_{8,2}$. Using the Buscher rules we have shown that the $5_2^2$ indeed couples to a 
mixed symmetry form but the two additional indices should not be thought of as form indices but appear more naturally as vector indices.
For this reason our notation includes superscripts in order to emphasize that these are indeed vector indices. It should be noted though that the 10D interpretation only makes sense in the presence of two isometry directions upon which no field depends.

Using S-duality we can determine the top-form coupling for the $5^2_3$-brane, which again will be 
written in terms of a contracted 8-form with two internal vector indices,
\beq\label{532WZterms}
\iota_y\iota_z C_{8}^{yz}.
\eeq
The definition of $C_8^2$ is found by S-dualizing eq. \eqref{B82definition}. At the linearized level this is
\[
\dd \iota_y\iota_z C_8^{yz} = \frac{\e^{-2\phi}}{|\tau|^3}\sqrt{\det(G_{kl})}\det(|\tau|G_{np} - C_{np})\ \hat\star\, \dd \tilde C^{yz}.
\]
We do not include more terms in this expression, since it is very sensitive to the RR sector, which we truncated away before S-duality. 

\subsection{Modified Bianchi identity}

In order to determine the modified Bianchi identities as a result of having a $5^2_2$ source, one considers the type IIB 
supergravity action in addition to the $5^2_2$ coupling
\beq\label{fullaction}
S = S_\text{NSNS} + \mu_{5_2^2}\int \iota_y\iota_z B_8^{yz},
\eeq
where the latter integral is six-dimensional over the worldvolume of the brane and where we have only included the NSNS
action
\be\label{bulkns}
S_\text{NSNS} = \int \dd^{10}x ~\e^{-2\phi}\biggl(\star {\cal R}+4\,\dd \phi\w\star\,\dd\phi -\sfrac 12 H\w\star\, H\biggl),
\ee
since we are interested in the coupling to the NSNS sector. In eq. (\ref{bulkns}), $\cal R$ is the curvature scalar. 

Varying the combined action $S$ is not straightforward, since they are essentially written in 
terms of different variables. 
The first step is to rewrite the NSNS action in terms of $B_8^2$, which turns out to be possible upon performing a 
redefinition of the NSNS fields inspired by generalized complex geometry\cite{Gualtieri}.
Let us recall that the above background fields may be collected 
in the so-called generalized metric, 
\be 
\cal\label{hB}  H = \left(\begin{array}{cc} G-BG^{-1}B  & BG^{-1} \\ 
-G^{-1}B & G^{-1} \end{array}\right),
\ee 
where $G$ and $B$ are the 10D NSNS metric and 2-form respectively.
However, in generalized geometry this is not the 
most general parametrization of the generalized metric. 
Indeed, a more general parametrization was used for example 
in Ref. \cite{aldazabal1} and reads as
\be\label{hgen}
 \cal H = \left(\begin{array}{cc} g-B g^{-1}B  & B g^{-1}+g\beta \\ 
-g^{-1}B-\beta g & g^{-1}-\beta g \beta \end{array}\right).
\ee
In the latter expressions $\beta=\beta^{mn}\partial_m\wedge \partial_n$ is a 2-vector, 
which appears naturally in generalized geometry, and $g$ and $B$ are 
a pair of different metric and 2-form than those that appear in eq. \eqref{hB}. This 
expression is the most general under the assumption{\footnote{We could easily raise this assumption but 
this is not essential for our present purposes. The parametrization without this assumption appears for example in
Ref. \cite{aldazabal1}.}} $B\beta=
\beta B=0$. 
In fact, for the case under study, it is enough to set $B=0$  
and determine the background in terms of $g$ and $\beta$ only
{\footnote{This was assumed in Ref. \cite{andriot} and it is enough for some simple cases 
like the one we discuss in this section. However, as pointed 
out in Ref. \cite{cj3}, it is not sufficient for more elaborate cases.}}.
Inspired by this reparametrisation let us write
\beq\label{redefinition}
G = (g^{-1} - \beta g \beta)^{-1},\qquad B = (g^{-1}(\beta)^{-1}g^{-1} - \beta)^{-1}.
\eeq
From this we calculate\cite{andriot}
\[
E^{-1} = (G + B)^{-1} = g^{-1} - \beta,
\]
which shows that the components of the field $\tilde B$, defined in eq. \eqref{BRBinternal} 
are identical to the components of the 2-vector $\beta$
along the isometry directions,
$$\tilde B^{mn}=\beta^{mn}.$$ As discussed above $\beta$ can be thought of as a 2-vector-valued scalar
\footnote{When referring to $\beta$ we will use 2-vector and 2-vector-valued scalar interchangeably.} 
with field strength 
\beq\label{Qdef}
Q_1^2 = \dd \beta,
\eeq
a 2-vector-valued 1-form. The magnetic dual of $\beta$ is a 2-vector-valued 8-form, i.e. it is essentially the 
mixed symmetry field $B_8^2$, at least at the linearized level. 
Hence we denote $B_8^2$ as $\beta_8^2$ from now on. The latter is defined as
\beq\label{Q9def}
Q_9^2 = \dd \beta_8^2 = \star \e^{-2\phi} \dd \beta.
\eeq
This equation serves also as a definition of the field strength $Q_9^2$.

As for the rewriting of the NSNS action, this was essentially done in Ref. \cite{andriot} (further approaches include \cite{Blumenhagen:2013aia,Andriot:2013xca}). 
The effective action is written in terms of the metric $g$ in the $\beta$-parametrization of
the generalized metric ${\cal H}$, the corresponding modified dilaton $\tilde\phi$ and the field strength (\ref{Qdef})
of the two-vector, 
\be 
S_\text{NSNS}= \int \dd^{10}x \sqrt{-g}~\e^{-2\tilde{\phi}}\biggl(\tilde {\cal R}+4|\dd \tilde{\phi}|^2 
-\sfrac 1{12} Q_M^{NR}Q^M_{NR}\biggl) + \int \dd(\cdots).
\ee 
The last term of this action is a total derivative which does not modify the equations of motion. 
Performing this variation, the modified Bianchi identity for $Q_1^2$ is determined to be
\beq\label{modifiedBI}
\dd (Q^{MN}_1\w g_{My}\dd y\w  g_{Nz} \dd z) = \star j_6^{5_2^2} ,
\eeq
where $\star j_6^{5_2^2} = \mu_{5_2^2}\delta_4$.

It is directly observed that the presence of a $5^2_2$ source turns on components of the field strength $Q$. 
This should be interpreted as a non-geometric flux in accord with the arguments of Ref. \cite{andriot}. Indeed, 
the whole point of rewriting the NSNS action in terms of the new variables is that, unlike the standard bulk 
action, it is well-defined for non-geometric backgrounds. This is expected in view of the fact that the standard 
NSNS action is not duality invariant. Then the interpretation is that the legitimate form of the bulk action depends 
on the duality frame of the background, i.e. the standard action is valid for geometric backgrounds and the 
rewritten one for non-geometric ones. Note that such a correspondence was also observed in the framework of 
matrix model compactifications in Ref. \cite{cj}.

Finally, concerning the $5_3^2$-brane, a set of S-dual statements hold. The role of $\tilde B$ is now played by 
$\tilde C$, also a 2-vector-valued scalar with field strength
\be 
P_1^2=\dd \tilde C.
\ee
Although a rewriting of the supergravity action for the RR sector is not known and generalised complex geometry 
is not sufficient to account for this, it is anticipated from S-duality that the corresponding Bianchi identity is 
\beq\label{modifiedBI}
\dd (P^{MN}_1\w g_{My}\dd y\w  g_{Nz} \dd z) = \star j_6^{5_3^2}.
\eeq
We will argue in section \ref{RRnongeometry} that this situation can be understood in the framework of extended generalised geometry. 
However, it should be kept in mind that a clear justification of this equation would be possible only after a 
successful reformulation of IIB supergravity in terms of the suitable variables. This is a task which deserves 
separate treatment and we will not perform it in the present paper.

\subsection{Relation to non-geometry}

The relation between exotic branes and non-geometric backgrounds
was first suggested in Ref. \cite{exotic,Hassler}, and can be understood as follows. From a lower-dimensional viewpoint 
(e.g. in three dimensions), supergravity scalar moduli 
undergo monodromies when they run around exotic point-particle 
states. 
These monodromies are essentially elements of the 
lower-dimensional U-duality 
group (e.g. $E_{8,8}(\Z)$ in three dimensions) which arises 
upon toroidal dimensional reduction of the higher-dimensional theory. 
However, from a higher-dimensional viewpoint these moduli are 
components of the supergravity fields (metric, NSNS and RR 
gauge potentials) and exotic states are codimension-2 
extended objects. Then the lower-dimensional monodromies 
are seen by a ``higher-dimensional observer" 
as a multivaluededness of the supergravity background. 
This phenomenon is one manifestation of what is usually termed non-geometry. 
The terminology stems from the fact that the background fields 
cannot be patched locally by standard geometric transition 
functions (diffeomorphisms for the metric, gauge transformations 
for the gauge potentials). In that sense they are globally ill-defined. 
In the brane picture, the background   
is mapped to a U-dual version when going around an exotic brane.

Given that  non-geometry was first discussed in the context of flux compactifications, it is useful to examine the relation 
between the compactification picture and the brane picture. This is possible due to the fact that branes may also 
be considered as supergravity solutions. This is of course true for D$p$-branes and NS5-branes, as well as for 
KKMs. Apart from these solutions, there also exist generalised KKMs \cite{gkkm}.
The background fields for the $5_2^2$-brane are given by
\bea 
\dd s^2&=&H(\dd r^2+r^2\dd \theta^2)+HK^{-1}(\dd y^2+\dd z^2)+(\dd x^{034567})^2, \nn\\ \label{522b}
e^{2\phi}&=&HK^{-1}, \\
B&=&-K^{-1}\theta \sigma \dd y\w \dd z,\nn
\eea
where 
$$
K=H^2+(\sigma\theta)^2,\qquad \sigma = R_yR_z,
$$ 
$H$ is a harmonic function and $R_y,R_z$ are the radii of the two circles of a 2-torus.
Clearly, the brane extends along 034567, the transverse corrdinates $yz$ correspond to a 2-torus and we use 
polar coordinates in the directions 12 in accord with Ref. \cite{exotic}. The multivaluededness of this solution
is manifest. Indeed, as $\theta \to \theta +2\pi $ all fields, the metric (its components in the $yz$ directions), the dilaton
and the NSNS potential (again the $yz$ components)  are not globally well-defined up to diffeomorphisms or gauge 
transformations. For example, since 
$$B_{yz}(r,\theta)=-\frac {\theta\sigma}{H^2+(\theta\sigma)^2},$$
we readily obtain
\be 
B_{yz}(r,\theta+2\pi)=-\frac {\theta\sigma+2\pi\sigma}{H^2+(\theta\sigma+2\pi\sigma)^2}
\ne B_{yz}(r,\theta)+\d B_{yz},
\ee 
for any gauge transformation $\d B$. The form of the NSNS potential is actually very suggestive. It is a more elaborate version of the usual illustrative example of a non-geometric background, where one starts with a 3-dimensional torus penetrated by NSNS flux. When two T-dualities are performed, the background fields cease to be globally well-defined. The result is what is usually termed a T-fold, because in order to patch fields on the manifold one has to use T-dualites as transition functions.

Using the background fields appearing in eq. (\ref{522b}), it is found that the $yz$ 
components of the generalized metric are given explicitly by the following 
$4\times 4$ matrix,
\be
{\cal H}_{5_2^2,(yz)} = \frac {1}{H}\left(\begin{array}{cccc} 1  & 0 & 0 & -{\s\theta} \\ 
0 &   1 & {\s\theta} & 0 \\ 0 & {\s\theta} &  K & 0 \\ 
-{\s\theta} & 0 & 0 &  K \end{array}\right),
\ee
where the parametrization (\ref{hB}) was used.
Note that there is nothing with a non-geometric flavour in the entries of this matrix. 
Non-geometry becomes apparent when the $B$ field components are written down.
However, the same generalized metric can be considered in the alternative parametrization of eq. (\ref{hgen}). 
In that case, the result is (see also Ref. \cite{geissbuhler2}),
\bea 
\dd s_{yz}^2&=&H^{-1} (\dd y^2+\dd z^2),\\
\beta &=& -\sigma\theta \partial_y\w \partial_z \quad \Rightarrow 
\quad \beta^{yz}=-\sigma\theta,\label{beta}
\eea 
where we exhibit only the components in the $yz$ directions, since the rest remains unaffected. 
Also, in this case $B=0$.
In generalized geometric terms this is a perfectly geometric 
expression under diffeomorphisms and $\beta$-gauge transformations. 
As such, this is understood as a geometrization of the 
apparently non-geometric background. This geometrization is based on the fact that in generalized geometry 
the structure group of the generalized bundle coincides with the T-duality group. Therefore T-dualities 
appear on equal footing with diffeomorphisms and gauge transformations of $B$ in this 
framework.
From eq. (\ref{Qdef}) it is found that the standard derivative of the $\beta$ components is
\be\label{qflux} Q_{\theta}^{\ yz}= \partial_{\theta}\beta^{yz},\ee
which gives the $Q$ flux, in accord with \cite{andriot}. 

In order to relate this solution to our previous discussion on the couplings, let us calculate the 8-form $\beta_8^2$ for this solution.
The relevant equation is \eqref{Q9def}, and it is important to note that in the present case the RR-corrections of 
that equation
are absent anyway, since all RR forms vanish for this background. 
A direct computation gives
\be \label{b82explicit}
\beta_8^{yz} = H\dd x^{034567}\w \dd y\w\dd z.
\ee
Accordingly, we obtain
\be 
\iota_y\iota_z \beta_8^{yz}=H\dd x^{034567}.
\ee
It is directly observed that this field is globally well-defined, much like its dual $\beta$. The following remarks 
are in order here. First of all, we observe that the expressions for $\beta$ and $\beta_8$ match with the 
corresponding ones for $B$ and $B_6$ of the NS5 solution,
\bea
B^{\text{NS5}}&=&\theta\sigma\dd y\w \dd z, \nn\\
B_6^{\text{NS5}}&=&H\dd x^{034567}. \nn
\eea
Secondly, we would like to stress that it does not make sense to calculate the magnetic dual, $B_6$, of the non-geometric $B$-field. Remember that $B_6$ is related to $B$ with a non-local expression and therefore one might get unexpected results for $B_6$ when $B$ is globally ill-defined.
For example, the authors of Ref. \cite{exotic} compute the standard dual $B_6$ of $B$ for the $5_2^2$-brane and they find it ill-behaved. 
We have shown here that this 
situation is not encountered when the proper field variables, $\beta$ and its magnetic dual, are used. 

Let us briefly mention that a second way to understand the above issues goes through the doubled 
formalism. This suggests the introduction of a dual set of 
coordinates, say $\check x_M$ (which may be understood as the Fourier 
duals for 
the winding numbers of the string in the same way that standard coordinates 
are the Fourier duals of its momenta). T-duality exchanges 
coordinates and dual coordinates much like it does for 
momenta and windings. From this ``generalized T-duality'' point of view there are no 
conventional Buscher rules. The $5_2^2$ solution, as 
double T-dual along $y$ and $z$ of the NS5, is just
\bea 
\check{\dd s^2}_{yz} &=& H\biggl(\dd \check y^2+\dd \check z^2\biggl),\\
\check B& =& -\sigma\theta \dd\check y\w \dd \check z. \label{checkB}
\eea 
These expressions are well-defined in the appropriate polarization of the doubled set of coordinates in the 
language of Ref. \cite{hulltdt}. We are not going to elaborate further on the treatment of the brane worldvolume actions 
in this framework in the present paper. However, it is useful to point out that the eqs. (\ref{beta}) and (\ref{checkB}) 
show the correspondence 
between the two approaches. When a globally ill-defined 
background is encountered one has to either transform to the 
bivector parametrization in generalized geometry or to 
the 2-form in dual coordinates in the doubled formalism. 
The components in both cases are exactly equal.

\subsection{S-duality and RR non-geometry}\label{RRnongeometry}

Up to now we focused our attention to the $5_2^2$-brane and discussed its relation to NSNS non-geometry and the Q flux. 
It is very interesting to note that S-duality mediates non-geometry to the RR sector as well. 
Indeed, the $5_3^2$-brane is S-dual to the $5_2^2$ one, which is a T-fold 
as discussed above. What is then the $5_3^2$-brane? In order 
to answer this question let us write down explicitly the 
corresponding supergravity solution \cite{gkkm}. It reads as 
\bea 
\dd s^2&=&(HK^{-1})^{\sfrac 12}(\dd r^2+r^2\dd \theta^2)
+(HK^{-1})^{\sfrac 12}(\dd y^2+\dd z^2)+(HK^{-1})^{-\sfrac 12}(\dd x^{034567})^2, \nn\\
e^{2\phi}&=&(HK^{-1})^{-1}, \\
C_2&=&-K^{-1}\theta\sigma \dd y\w \dd z.\nn
\eea
It is immediately observed that this is also a non-geometric 
background. The main focus in this case should be on the 
fact that a globally ill-defined RR gauge potential
 is encountered. This fact barely needs any further explanation; 
it is just the S-dual version of the $Q$ flux non-geometry. 
The $B$ is exchanged with $C_2$ and now for the latter holds that
\be 
C_{yz}(r,\theta+2\pi)
\ne C_{yz}(r,\theta)+\d C_{yz},
\ee 
where a gauge transformation for $C_2$ is 
$$\d C_2=\dd \l_1.$$

The global issues of this solution are expected to be resolved in a similar 
way as previously. However, one has to implement the RR sector into the discussion and therefore 
generalized geometry is not sufficient anymore. Extensions of generalized geometry to account for the RR sector 
and the full U-duality group were suggested in Refs. \cite{egg1,Pacheco:2008ps} (see also \cite{BermanExtGeo,Waldram,Park:2013gaj,Cederwall:2013naa,Andriot:2013xca}). In this exceptional (or extended) generalized geometry 
setting, the structure group of 
the generalized bundle is extended from the T-duality group to the U-duality one. 
In such terms, we have to consider 
a new 2-vector 
$$\gamma=\gamma^{mn}\partial_m\wedge\partial_n$$
and the associated $\gamma$-transformations. Such quantities were 
 considered in Ref. \cite{aldazabalRR1} and studied further in Ref. \cite{rrng} using 
 group-theoretical techniques. In the context of what we have considered in this paper, the components of this 
 2-vector are equal to the ones of $\tilde C$, defined in (\ref{tildeC}) and being the S-dual of $\tilde B$, i.e.
 $$\g^{mn}=\tilde C^{mn}.$$
 
The flux associated to this solution is given by the 
derivative of $\g$ with respect to $\theta$ and we denote it 
by $P$, as in Refs. \cite{aldazabalRR1,rrng},
\be P_{\theta}^{ \ yz}=\partial_{\theta}\g^{yz}. \ee
This is the RR analogue of the Q flux. They have the same index structure and 
number of components. As such, $P$ is to
 $F_3$  flux what $Q$ is to $H_3$ flux. This essentially 
augments the chain of fluxes as
\be 
F_{abc} \overset{S}\longleftrightarrow H_{abc} \overset{T_a}\longleftrightarrow f^a_{\ bc} \overset{T_b}
\longleftrightarrow Q^{ab}_{\ \ c}
 \overset{S}\longleftrightarrow P^{ab}_{\ \ c}.
\ee 
In order to avoid confusion, let us note that the middle 
entry $f$ refers here to the type IIA theory. 

The above discussion shows that the $5_3^2$-brane is a U-fold associated to structures that 
appeared before in the context of exceptional generalised geometry.

\section{Discussion}\label{summary}

\paragraph{Summary of results.} Non-standard branes are very interesting BPS states of string theory originating from string dualities. They are generically heavier than the standard D$p$-branes, exhibiting a wide range of power dependencies on the inverse string coupling as well as a number of NUT type transverse directions. What is more, they induce non-trivial monodromies around them, an effect that leads to interesting consequences; in particular they correspond to some of the so-called non-geometric string backgrounds.
In the present paper we studied some aspects of the physics of such non-standard branes. 
Focusing on the type IIB superstring and in particular on the fivebranes of this theory, we first revisited the worldvolume actions of the standard fivebranes (the NS5-brane and the Kaluza-Klein monopole) and expressed them in a compact form. Subsequently, we derived and applied the appropriate set of duality rules (including T and S-duality) with the aim of determining the analogs of the DBI action for non-standard fivebranes. 
These actions appear in section \ref{exoticDBI} and they can be useful in a further study of the dynamics of such branes. A very interesting question concerns the couplings of these branes and their role as sources in the type IIB theory. In section \ref{exoticWZ} we clarified these couplings up to contributions from the RR sector. In particular, we showed that one of the exotic fivebranes, the $5_2^2$, couples to an exotic magnetic dual of the Kalb-Ramond field, in accord with previous expectations. Together with a rewriting of the bulk NSNS action, we utilized this coupling to derive a modified Bianchi identity. 
This leads to the result that the $5_2^2$-brane is a source of non-geometric Q flux. The analogous treatment for the second exotic fivebrane of type IIB leads to its coupling to an exotic dual of the RR 2-form and a corresponding modified Bianchi identity. The latter renders the $5^2_3$-brane a source of non-geometric RR flux, which we call P. Finally, we examined the connection among the above brane picture and the more familiar picture of flux compactifications. As advocated before in the literature, non-standard branes correspond to U-folds, 
the latter being spaces which allow patching of fields with U-duality transformations. We examined in some detail this relation from the point of view of generalized complex geometry for the $5_2^2$-brane. Moreover, we discussed a similar treatment within a broader context of extended generalized geometry for its S-dual $5_3^2$-brane and advocated for a relation of the latter to non-geometric RR flux. The relation between the brane picture and the flux compactification picture may be summarized in the following diagram,

\be
\begin{array}{ccccccccc}
F_{abc}&\overset{S}\longleftrightarrow& 
H_{abc} &\overset{T_a}\longleftrightarrow &f^a_{\ bc} &\overset{T_b}\longleftrightarrow &Q^{ab}_{\ \ c} & \overset{S}
\longleftrightarrow &P^{ab}_{\ \ c}\\
\biggl\updownarrow& & \biggl\updownarrow  && \biggl\updownarrow && \biggl\updownarrow& & \biggl\updownarrow
\\
\text{D5}&\overset{S}\longleftrightarrow&\text{NS5}& \overset{T}\longleftrightarrow& \text{KKM} & \overset{T}\longleftrightarrow 
& 5_2^2& \overset{S}\longleftrightarrow &5_3^2
\end{array}\nn\ee
The upper row depicts the compactification picture with the associated fluxes in each duality frame. The lower row depicts the fivebranes that we studied and their relation through dualities. Most importantly, the vertical arrows connect the corresponding entries of each row in the sense that each brane is a source of each type of flux.

\paragraph{Limitations.}
There exist a number of limitations of our results that should be well kept in mind.
First of all, the gauge completion of the leading 
coupling in the WZ action for the exotic branes is not a trivial task. In order to perform it, one should know 
how the corresponding mixed symmetry fields appear in the supergravity actions. Although we took some steps in this 
direction, one would need a full
reformulation of the standard type II supergravities in terms of these fields, which is still lacking. 
Secondly, the non-standard branes that we considered in this paper do not have  finite energy density as single objects in flat space. This is not surprising in view of similar features 
associated to more standard branes. In particular, the D7-brane (also a co-dimension-2 object) exhibits the same 
behaviour \cite{Bergshoeff:1996ui}. In addition, branes such as the KKM have special NUT-like direction that require circular isometry and cannot be made non-compact. Non-standard branes 
essentially carry both problematic aspects as single objects, being at the same time co-dimension-2 and having 
several special directions as well{\footnote{We thank Eric Bergshoeff for pointing this out to us.}}. 
Therefore, a more consistent study of such branes should involve some additional mechanism, either 
involving angular momentum (as in Ref. \cite{exotic}) or a generalized Myers effect (which demands knowledge of 
non-abelian couplings of several coincident branes). Moreover, it should be mentioned that our derivations reside fully 
on duality rules. In this sense the worldvolume actions for all but the D$p$-branes cannot be directly associated to 
a worldsheet computation and therefore corrections are harder to determine. 
Finally, we would like to remind that branes associated to the so-called $R$ flux were not considered here (see 
Ref. \cite{Hassler} for a discussion on a possible description of such branes). Although it remains unclear to us how 
to treat such branes with the methods used in this paper, the experience from flux compactifications points to 
co-dimension-1 branes (recall that D8 is co-dimension-1 too).
These ``domain wall branes'' were classified in Ref. \cite{domainwalls} and it would be interesting to examine 
potential relations to R-type non-geometric fluxes.

\paragraph{Acknowledgements.}
Useful discussions with E. Bergshoeff, L. Jonke, O. Lechtenfeld and D. L\"ust are gratefully acknowledged. This work was supported in part by the German Research Foundation (DFG) within the Cluster of Excellence ``QUEST" and the Riemann Center for Geometry and Physics.

\appendix
\section{Conventions and notation}\label{notation}
We use the democratic formulation of type IIB supergravity
\beq\label{typeIIaction}
S = \int \e^{-2\phi} \left(\star {\cal R} + 4 \dd \phi\w\star\dd \phi - \frac12  H\w \star H - \frac14\e^{2\phi}  F\w \star F \right),
\eeq
where ${\cal R}$ is the Ricci scalar associated with the metric $G_{MN}$ with mostly plus signature, $\phi$ is the dilaton, $H$ is the NSNS three form field strength and $F = \dd {\cal C} - H\w {\cal C}$ are the RR field strengths. We use polyform notation for the RR potentials and field strengths such that ${\cal C}$ is the formal sum
\[
{\cal C} = C_0 + C_2 + C_4 + C_6 + C_8.
\]
The self-duality relation for the field strengths $\star F_p = (-1)^{p(p-1)/2} F_{10-p}$ must be imposed by hand at the level of equations of motion.
A general $p$-form $\omega_p$ is decomposed in the basis forms $\dd x^M$ as
\[
\omega_p = \frac{1}{p!}\omega_{M_1\cdots M_p} \dd x^{M_1} \w \cdots \w \dd x^{M_p}
\]
and our Hodge star is defined such that
\[
\omega_p \w \star\omega_p = \frac{1}{p!} \omega_{M_1\cdots M_p} \omega_{M'_1\cdots M'_p} G^{M_1M'_1}\cdots G^{M_p M'_p}\ \sqrt{-\det G}\ \dd x^0\w \cdots \w\dd x^{9},
\]
which means that
\[
\star \dd x^0\w \cdots \w \dd x^p = \sqrt{-G}\  \dd x^{p+1} \w \cdots \w \dd x^9.
\]
For most $p$-forms we will explicitly denote their degree with a subscript $p$, but for common forms such as the 2-form $B$, we will omit the subscript. When we write out the components of forms, in order not to clutter our notation, we will also omit writing the degree subscript. In the paper we encounter objects that carry vector indices. In that case the number of vector indices are denoted with a superscript. The superscript is also omitted for commonly used objects or when the components are written out explicitly. For the above conventions the WZ terms of D$p$-branes take the form
\[
\mu_p \int \e^{-{\cal F}} {\cal C}|_{p+1},
\]
and the T-duality rules agree with those of Myers\cite{Myers:1999ps}, except for a flipped sign of $B$. Here we have used the symbol $|_p$ to denote that only the $p$ form of the corresponding polyform should be considered.

In the paper we use the following KK decomposition of the metric
\beq\label{KKansatz}
\dd s^2 = \hat G_{\mu\nu} \dd x^\mu\dd x^\nu + G_{mn} \eta^m \eta^n,
\eeq
where capital Latin indices are ten-dimensional, lower case Latin indices are $d$-dimensional where $d$ is the number of Killing isometries and lower case Greek indices are $(10-d)$-dimensional. The $d$ 1-forms $\eta^m$ are such that $k^M_m \eta_M^n = \delta^n_m$ and $k^M_m$ are components of the Killing vector associated to the $m$-th Killing direction. In fact $k_m = \partial_m$ and $\eta^m = \dd x^m + A^m$, where $A^m = A^m_\mu \dd x^\mu$ are the KK 1-forms. The contraction of a $p$-form $\omega_p$ with the Killing vector $\partial_m$ is denoted by
\[
\iota_m \omega_p = \frac{1}{(p-1)!} \omega_{mM_1\cdots M_{p-1}} \dd x^{M_1}\w\cdots\w \dd x^{M_{p-1}}.
\]
For the KK ansatz \eqref{KKansatz} the Hodge star has a simple decomposition, for example when $d=1$ we get for a $p$-form $\omega_p = \bar\omega_p + \omega_{p-1}\w \eta^z$ (we let $z$ label the isometry direction, the 9-th coordinate,)
\beq
\star \bar\omega_p &=& \sqrt{\det(G_{mn})}\ \hat\star\bar\omega_p \w \eta^z\notag\\
\star (\omega_{p-1} \w \eta^z) &=& (-1)^p\sqrt{\det(G_{mn})}\ G^{zz}\ \hat\star\omega_{p-1}, \notag
\eeq
where $\hat\star$ is the reduced Hodge star operator associated with the metric $\hat G$.

\section{Gauge transformations and S-duality in type IIB}\label{gaugetrafo}
Let us consider type IIB supergravity. It contains the Kalb-Ramond potential $B$ with field strength
\be \dd H=0\Rightarrow H=\dd B \ee 
and the Ramond-Ramond potentials $C_0,C_2,C_4,C_6,C_8$. 
The equations of the theory may be written as Bianchi identities.
In the conventions we use, the equations for the field strengths take the form
\bea 
\dd F_1&=&0, \nn\\
\dd F_3&=&-F_1\wedge H, \nn\\
\dd F_5&=&-F_3\wedge H, \nn\\
\dd F_7&=& -F_5\wedge H, \nn\\
\dd H_7&=&-F_1\wedge F_7+F_3\wedge F_5. \nn
\eea 
In addition, these field strengths must be gauge invariant. This property allows us to determine the gauge 
transformation rules of the corresponding gauge potentials. 
In the simplest case of the Kalb-Ramond field, gauge invariance gives
\be \label{deltaB}\delta H=0 \Rightarrow \d B=\dd\L_1, \ee
where $\L_1$ is the gauge transformation parameter and it is an 1-form. In general, we denote gauge parameters as
$\L_p$ and $\l_p$, where the capital ones are reserved for $B$ and its magnetic dual and lower case ones are used for the 
RR potentials. The index $p$ declares the degree of the form.

 The field strength of $C_0$ is 
 \be F_1=\dd C_0 \ee 
 and gauge invariance gives
 \be\label{deltaC0} \d F_1=0 \Rightarrow \d C_0=0. \ee
 The equation for the field strength $F_3$ is solved by
 \be F_3=\dd C_2-C_0H, \ee
and therefore
\be \label{deltaC2}\d F_3=0 \Rightarrow \dd(\d C_2)=0 \Rightarrow \d C_2=d\l_1. \ee
Moreover, the equation for $F_5$ may be solved as
\be F_5=\dd C_4-C_2\wedge H, \ee
which means that 
\be\label{deltaC4}\d F_5=0 \Rightarrow \dd(\d C_4) - \dd\l_1\wedge H=0 \Rightarrow \d C_4 = \dd\l_3 + \dd\l_1\wedge B. \ee
Of course, there is no unique way to solve the corresponding identity. We could have solved it instead as
\be \label{F5b}F_5=\dd C'_4-\sfrac 12 C_2\wedge H+\sfrac 12 B\wedge \dd C_2. \ee
This is for example the definition used in Ref. \cite{polchinski}. In this case we obtain
\be \label{deltaC4b}\d F_5=0 \Rightarrow \d C'_4=d\l_3 +\sfrac 12 \dd\l_1\wedge B-\sfrac 12 \dd\L_1\wedge C_2. \ee
This is in agreement with the conventions of Ref. \cite{lozano1} and moreover it has a flavour of S-duality, 
unlike the previous expression. 
Of course, the choices are equivalent, since they both solve the same Bianchi identity. 
In this paper we are going to use the former choice.

Finally, we need the gauge transformation of $C_6$, which is the form that couples (electrically) to the D5-brane.
Its curvature may be defined as
\be\label{F7a} F_7 = \dd C_6 -C_4\wedge H \ee
or
\be\label{F7b} F_7 = \dd C'_6 -C'_4\wedge H + \sfrac 14 B\wedge B\wedge \dd C_2, \ee
which are both consistent with the corresponding Bianchi identity, the latter being consistent when $F_5$ is defined through 
the expression \eqref{F5b}. As before, we are using the former expression. 
This leads to the following gauge transformation,
\be \label{delta6b}
\d C_6 = \dd \l_5 + \dd \l_3\w B + \frac{1}{2}\dd \l_1 \w B \w B.
\ee

What remains is the gauge transformation of the field $B_6$, the magnetic dual of the Kalb-Ramond potential $B$. 
This transformation is necessary in several instances in the main text. 
 It can be determined from the equation of motion for 
$H$ in the type IIB superstring, which translates into a Bianchi identity for $H_7$
\be \dd H_7+F_1\wedge F_7-F_3\wedge F_5=0. \ee
This equation is solved by the following field strength,
\be 
\label{h7}
H_7=\dd B_6- C_4\wedge \dd C_2-\sfrac 12 C_2\wedge C_2\wedge H -C_0F_7.
\ee 
Subsequently, imposing the gauge invariance of $H_7$,
$$ \d H_7=0,$$
we determine the gauge transformation of the $B_6$ to be
\be\label{deltaB6} \d B_6=\dd\L_5+ \dd\l_3\wedge C_2
+\dd\l_1\wedge B\wedge C_2, \ee
where $\L_5$ is a 5-form gauge parameter.

For completeness we list here the S-duality rules using the above conventions,
however we omit the S-duality relation for $C_8$ since we do not use it in this
paper. These relations read as follows,
\beq
G&\sdual& |\tau|G,\notag\\
\tau&\sdual& -\frac{1}{\tau},\notag\\
C_2 &\sdual& B,\notag\\
B   &\sdual& -C_2,\notag\\
C_4 &\sdual& C_4 - C_2\w B,\notag\\
C_6 &\sdual& -B_6 + \frac{1}{2}B\w C_2\w C_2,\notag\\
B_6 &\sdual& C_6 - \frac{1}{2} C_2 \w B \w B,
\eeq
where $\tau = C_0 + i\e^{-\phi}$ is the type IIB axio-dilaton. Using these rules we can easily write down the mapping of 
the polyform ${\cal C}$ under S-duality,
\[
{\cal C} \sdual - {\cal B} =  -\frac{C_0}{|\tau|^2} + B + \left(C_4 - C_2\w B\right)+ \left(-B_6 + \frac{1}{2}B\w C_2\w C_2\right)+
\tilde C_8
\]
where $\tilde C_8$ is the S-dual of $C_8$. This should be inserted in to eq. \eqref{wzns5b} to give the correct WZ terms 
for the NS5-brane. One can then directly check the gauge invariance of eq. \eqref{wzns5b} using the gauge transformations 
(\ref{deltaB},\ref{deltaC0},\ref{deltaC2},\ref{deltaC4},\ref{deltaB6}).

\section{Reduced action and magnetic duals}\label{appreduced}
For the ansatz (\ref{twodmetric},\ref{Bfield})
\beq
\dd s^2 &=& \hat{G}_{\mu\nu} \dd x^\mu\dd x^\nu + G_{mn}\eta^m\eta^n,\notag\\
B &=& \hat{B} + (\eta^m - \frac{1}{2} A^m)\w \theta_m + \frac12 B_{mn}\eta^m\w\eta^n,\notag\\
\e^{2\phi} &=& \sqrt{\det(G_{mn})}\ \e^{2\hat\phi},
\eeq
the NSNS action
\[
S_{\text{NSNS}} = \int \e^{-2\phi}\ \left(\star\,{\cal R} + 4\dd \phi \w\star\,\dd\phi - \frac12  H\w\star\, H\right)
\]
reduces to\cite{ms,KaloperMyers,hullortin}
\beq
S_\text{NSNS} &=& \text{Vol}_d \int  \e^{-2\hat\phi} \left( \hat\star\,\hat{\cal R} + 4\dd \hat\phi\w \hat\star\,\dd \hat\phi - \frac12\hat H \w\hat\star\,\hat H\right.\notag\\
	&& \left.+ \frac18 \tr(\dd M \w\hat\star\,\dd M^{-1}) - \frac12 M_{IJ}\dd{\cal A}^I\w\hat\star\,\dd{\cal A}^J  \right)\label{NSNSreduced} 
\eeq
where $d$ denotes the dimension of the space which we reduce on,
\[
M_{IJ} = \left(\begin{array}{cc} G_{mn}- B_{mm'}G^{m'n'}B_{n'n}& B_{mm'}G^{m'n}\\ - G^{mm'}B_{m'n} &G^{mn}\end{array} \right)
\]
is the scalar moduli matrix, $I,J$ are $O(d,d)$ indices and 
\[
{\cal A}^I = \left(\begin{array}{c} A^m \\ -\theta_m \end{array}\right).
\]
Then $\hat H$ can be neatly expressed as
\[
\hat H = \dd \hat B - \frac12 L_{IJ}{\cal A}^I \w \dd {\cal A}^J,\qquad  L_{IJ} = \left(\begin{array}{cc} 0&1\\1&0 \end{array} \right).
\]
Variation of the action \eqref{NSNSreduced} with respect to ${\cal A}$ gives the equations of motion
\beq
\dd \left( \e^{-2\hat\phi} M_{IJ}\ \hat\star\ \dd {\cal A}^I\right) = -L_{IJ} \dd {\cal A}^I\w\e^{-2\hat\phi}\ \hat\star\ \hat H.
\eeq 
Here we have used the $\hat B$ equation of motion
\beq
\dd \left( \e^{-2\hat\phi} \hat\star \hat H\right) = 0.
\eeq
The magnetic dual of ${\cal A}$ can now be consistently defined as
\beq
\dd {\cal A}_{J,7-d} = \e^{-2\hat\phi} M_{IJ}\ \hat\star\ \dd {\cal A}^I + L_{IJ} {\cal A}^I\w \e^{-2\hat\phi}\ \hat\star\ \hat H.
\eeq
We are interested in the nine dimensional case where we encounter the magnetic dual of $A^m$ which couples to the KKM. In that case $M_{IJ}$ takes a diagonal form and we get
\beq
\dd A_6^z &=& \e^{-2\hat\phi} G_{zz}\ \hat\star\ \dd A^z - \theta_z\w\iota_z \dd B_6,\notag\\
\dd \theta_{z,6} &=& \e^{-2\hat\phi} G^{zz}\ \hat\star\ \dd \theta_z - A^z\w \iota_z\dd B_6,\notag
\eeq
where $B_6$ is the ten-dimensional magnetic dual of $B$. We can directly relate the expression for $\dd \theta_{z,6}$ to the ten-dimensional field $B_6$
\[
\dd \theta_{z,6}  = \dd(1-P_z)B_6.
\]
The expression for $\dd A_6^z$ should also be related directly to a ten-dimensional field, but this will not be a  6-form but the 7-form $A_7^z$ which is the ten-dimensional magnetic dual of $A^z$,
\beq\label{A7def}
\dd A_6^z = \dd \iota_z A_7^z.
\eeq


\end{document}